\documentclass[aps,prl,nofootinbib,reprint]{revtex4-1}

\pdfoutput=1
\usepackage{graphicx}

\usepackage{amsmath,amssymb,mathrsfs}

\usepackage[bookmarks=false]{hyperref} 
\hypersetup{pdfstartview=FitH,pdfhighlight=/O,colorlinks=false}

\begin{document}

\title{Mechanical laws of the Rindler horizon}

\author{Eugenio~Bianchi}
\affiliation{Perimeter Institute for Theoretical Physics\\
31 Caroline St.N, Waterloo ON, N2J 2Y5, Canada\\
\href{mailto:ebianchi@perimeterinstitute.ca}{\tt ebianchi@perimeterinstitute.ca}}

\author{Alejandro~Satz}
\affiliation{Maryland Center for Fundamental Physics\\
Department of Physics, University of Maryland\\
College Park, MD 20742-4111, USA\\
\href{mailto:alesatz@umd.edu}{\tt alesatz@umd.edu}}


\begin{abstract}

Gravitational perturbations of flat Minkowski space make the Rindler horizon dynamical: the horizon satisfies mechanical laws analogous to the ones followed by black holes. We describe the gravitational perturbation of Minkowski space using perturbative field-theoretical methods. The change in the area of the Rindler horizon is described in terms of the deflection of light rays by the gravitational field. The difference between the area of the perturbed and the unperturbed horizon is related to the energy of matter crossing the horizon. We derive consistency conditions for the validity of our approximations, and compare our results to similar ones present in the literature. Finally, we discuss how this setting can be used in perturbative quantum gravity to extend the classical mechanical laws to thermodynamic laws, with the entanglement of field modes across the Rindler horizon providing a notion of thermodynamic entropy.
\end{abstract}

\maketitle

\section{I. Introduction} \vspace{-1em}

The intimate relationship between the geometrical properties of spacetime horizons and the laws of thermodynamics has been known for four decades and plays still a key role in the quest for a quantum theory of gravity. Its history originates with the formulation by Bardeen, Carter and Hawking \cite{Bardeen:1973gs,Hawking:1972hy,1979grec.conf..294C} of the Laws of Black Hole Mechanics. These are summarized as: 0) the constancy of the surface gravity $\kappa$ of a stationary black hole over the event horizon; 1) the mass difference between two neighboring stationary black holes satisfying $\delta M = (8\pi G)^{-1}\kappa\,\delta A+\delta W$, where $A$ is the event horizon area and $\delta W$ a work term; 2) the non-decreasing nature of the horizon area, and 3) the impossibility to obtain $\kappa=0$. Together with Bekenstein's conjectured identification of black hole event horizon area with entropy \cite{Bekenstein:1973ur}, these results uncovered a powerful analogy between the two seemingly unrelated fields of general relativity and thermodynamics. This was fully vindicated when Hawking derived the existence of black hole thermal radiation \cite{Hawking:1974rv,Hawking:1974sw}, showing that the relation $\kappa = 2\pi T$ is an identity rather than a mere analogy. 

It is well known that other horizons not tied to black holes, such as the de Sitter horizon and the Rindler horizon in Minkowski space, also exhibit thermal properties when quantum theory is taken into account \cite{Unruh:1976db,Gibbons:1977mu}. It is therefore natural to presume that they also satisfy appropriate mechanical laws which correspond to thermodynamical laws in the same way as the black hole mechanics laws do. We are especially interested in this question for the Rindler horizon, since by the equivalence principle all causal horizons are similar to the Rindler horizon in a sufficiently small region.

We should not expect a perfect correspondence with the laws of black hole mechanics. One obvious difference is that the area of the Rindler horizon is infinite -- and hence, presumably its entropy. However, this is not a big concern as one can analyze the area of a localized subregion of the horizon and see if it satisfies laws (1) and (2). A second difference is in the role of the surface gravity $\kappa$ that is now observer-dependent. In the mechanical laws of the Rindler horizon,  $\kappa$ is identified with the acceleration of a Rindler observer, and $\delta M$ with the energy that crosses the horizon as measured by the same observer. On the other hand, the change $\delta A$ in the area of the Rindler horizon is independent from the acceleration $\kappa$.

The mechanical laws of the Rindler horizon were first discussed by Jacobson in \cite{Jacobson:1999mi} and studied by Jacobson and Parentani in \cite{Jacobson:2003wv}. Further discussion of the validity of these laws can be found in \cite{Amsel:2007mh}, in \cite{Jacobson:1995ab,Eling:2006aw,Brustein:2009hy,Parikh:2009qs,Padmanabhan:2009ry,Chirco:2010xx,Yokokura:2011za,Guedens:2011dy,Padmanabhan:2009vy}, in \cite{Ghosh:2011fc,Frodden:2011eb}, and in \cite{Wall:2010cj,Wall:2011hj}. The methods employed in these works follow closely the ones adopted in the analysis of black hole mechanics \cite{Wald:1995yp}.

The motivation for this paper is twofold. First, we want to provide a pedagogical treatment of horizon mechanics that does not presuppose previous knowledge of general relativity or of the differential geometry of null geodesic congruences. This is possible thanks to the fact that we study small perturbations about a flat Minkowski background. In doing this, we extend the field-theorist's treatment of ``Feynman Lectures on Gravitation'' \cite{Feynman:1996kb} to the realm of horizon mechanics.

Our second motivation is to use these newly developed field-theoretical tools to address open questions in horizon physics. At the classical level, the description of the mechanics of the Rindler horizon in terms of a gravitational field $h_{\mu\nu}(x)$ allows to develop a perturbative scheme with advantages over the geometric description, similar to the ones encountered in gravitational wave physics \cite{Maggiore:1900zz}. For instance, the methods we introduce provide a perturbative scheme for the study of approximate boost Killing vectors \cite{Guedens:2011dy}. Moreover, the methods of perturbative quantum gravity \cite{Donoghue:1995cz} can be immediately employed to extend the classical laws to quantum laws and provide an effective field theory description of the quantum Rindler horizon. Conjoined with standard results about the Unruh effect \cite{Unruh:1976db} and recent developments in the analysis of entanglement entropy in quantum field theory \cite{Bianchi:2012br}, this leads to a derivation of the thermodynamic laws of the Rindler horizon.

The perturbative dynamics of the Rindler horizon is studied in section II. The first and the second laws are derived in sections III and IV. Sections V-VII provide a physical interpretation and an analysis of the validity of these laws. In section VIII we investigate quantum aspects of horizon mechanics.

Throughout the paper spacetime is four-dimensional and flat, and we use Cartesian coordinates $x^{\mu}=(t,\vec{x})=(t,x,y_1,y_2)$. The Minkowski metric $\eta_{\mu\nu}$ has signature $(-+++)$, and indices are always raised and lowered using $\eta_{\mu\nu}$. Everywhere except in section VI, we use units  $c=1$ and $\hbar=1$.

\section{II. Area of the Rindler horizon in perturbative gravity} \vspace{-1em}

In this section we derive a formula for the area of the Rindler horizon in the presence of a gravitational perturbation $h_{\mu\nu}$ on flat Minkowski space. We adopt a ``particle physicist's'' point of view \cite{Feynman:1996kb,Weinberg:gravitation} minimizing the geometrical interpretation of the gravitational field. 

\subsection{The gravitational field} \vspace{-1em}
Weak gravitational phenomena in flat Minkowski space can be described perturbatively in terms of a gravitational field $h_{\mu\nu}(x)$. In the following we partially fix the gauge symmetry of this field by imposing the harmonic gauge
\begin{equation}
\partial_\mu h^{\mu}_\nu-\frac{1}{2}\partial_\nu h^\mu_\mu=0\,.\label{eq:harmonic}
\end{equation} 
At the lowest order in Newton's coupling constant $G$, the gravitational field satisfies the wave equation
\begin{equation}\label{fieldequation}
\Box h_{\mu\nu}=-\sqrt{8\pi G}\,\big(T_{\mu\nu}-\frac{1}{2}\eta_{\mu\nu} T_\rho^\rho\big)\;,
\end{equation}
where $T_{\mu\nu}(x)$ is the energy-momentum tensor of matter and of the gravitational field,
\begin{equation}
T_{\mu\nu}\;=\;T_{\mu\nu}^{\text{matt}}+T_{\mu\nu}^{\text{grav}}\,,
\end{equation}
with $T_{\mu\nu}^{\text{grav}}\sim(\partial h)^2$, up to corrections of higher order in $\sqrt{8\pi G}$. In the same approximation, the gravitational field is given by 
\begin{equation}
h_{\mu\nu}(x)=h_{\mu\nu}^{\text{rad}}(x)+h_{\mu\nu}^{\text{source}}(x)\,.
\end{equation}
The first term is a radiative contribution consisting purely of gravitational waves, $\Box h_{\mu\nu}^{\text{rad}}=0$. The second term is determined by the total energy-momentum source
\begin{align}
&h_{\mu\nu}^{\text{source}}(x)=\\
&\quad=-\sqrt{8\pi G}\int \mathrm{d}^4 x'\,G_R(x,x')\,\big(T_{\mu\nu}(x')-\frac{1}{2}\eta_{\mu\nu} T_\rho^\rho(x')\big)\,,\nonumber
\end{align}
where $G_R(x,x')$ is the retarded Green function coding the fact that there is no incoming radiation from past null infinity (Kirchoff-Sommerfeld boundary condition),
\begin{align}
G_R(x,x')=&\frac{1}{2\pi}\delta(\|x-x'\|^2)\theta(t-t')\\
=&-\frac{1}{4\pi} \frac{1}{|\vec{x}-\vec{x}'|}\delta(t-t'-|\vec{x}-\vec{x}'|)\,.
\end{align}
This field-theoretical description of gravity is related to the geometric one proper to general relativity by the identification of the curved space-time metric $g_{\mu\nu}(x)$ with the perturbed Minkowski metric
\begin{equation}
g_{\mu\nu}(x)=\eta_{\mu\nu}+\sqrt{32\pi G}\, h_{\mu\nu}(x)\;.
\end{equation}
These two languages are equivalent and complementary for weak gravitational phenomena \cite{Maggiore:1900zz}.

\subsection{Deflection of light by the gravitational field} \vspace{-1em}
In the presence of a gravitational field, light rays follow a path $x^\mu(v)$ that extremizes the action $I=\int L \,\mathrm{d}v$ with  Lagrangian
\begin{equation}
L=\frac{1}{2}\eta_{\mu\nu}\dot{x}^\mu(v)\dot{x}^\nu(v)+\sqrt{8\pi G} h_{\mu\nu}(x(v))\,\dot{x}^\mu(v)\dot{x}^\nu(v)\,.
\end{equation}
The equations of motion are 
\begin{align}
\ddot{x}^\mu(v)=&\sqrt{8\pi G}\big(\partial^\mu h_{\rho\sigma}(x(v))\dot{x}^\rho \dot{x}^\sigma-2\partial_\rho h^\mu_\sigma(x(v))\dot{x}^\rho \dot{x}^\sigma \nonumber\\
&\qquad\quad-2h^{\mu}_\nu(x(v)) \ddot{x}^\nu\big)\,,
\end{align}
and the initial condition $L(v_0)=0$ for light rays is assumed. The free solution for $G=0$ is $\ddot{x}_0^\mu(v)=0$. We consider a beam of light rays starting at $t=0$ and $x=0$ with direction $k^\mu=(1,1,0,0)$, so that, unperturbed by the gravitational field, they follow the straight path
\begin{equation}
x_0^\mu(v)=y^\mu+v k^\mu\,,\label{eq:unperturbed}
\end{equation}
with $y^\mu=(0,0,y_1,y_2)$ and $v = (x+t)/2$. The light rays can be thought of as indexed by the value of their initial condition $(y_1,y_2)$. 

At the lowest order in $\sqrt{8\pi G}$, the trajectory $x^\mu(v)$ of the perturbed rays can be described in terms of a small deflection $\xi^\mu(v)$ from the straight path,
\begin{equation}
x^\mu(v)\,=\,x_0^\mu(v)+\xi^\mu(v)\,.\label{eq:path}
\end{equation}
We will work to the first order in both $\sqrt{G}$ and $\xi^\mu$.
For the deflection we find the equation of motion
\begin{equation}
\ddot{\xi}^\mu=\sqrt{8\pi G}\big(\partial^\mu h_{\rho\sigma}(x_0(v))-2 \partial_\rho h^\mu_\sigma(x_0(v)) \big)k^\rho k^\sigma\;.\label{eq:geo}
\end{equation}
Given a gravitational field $h_{\mu\nu}$, this equation can be integrated and the light path determined completely. Note that in principle, a solution to this equation could lead to large $\xi$, so we will need an additional consistency condition (to be discussed in Section VII) to ensure the validity of the approximation.

Notice also that from (\ref{eq:geo}) it follows that $k_\mu \ddot{\xi}^\mu=-\sqrt{8\pi G}\dot{h}_{\mu\nu}(x(v))k^\mu k^\nu$, so that if $\dot{\xi}^\mu(v_0)=0$ in a region where $h_{\mu\nu}(v_0)=0$, then 
\begin{equation}
k_\mu \dot{\xi}^\mu(v)=-\sqrt{8\pi G}h_{\mu\nu}(x(v))k^\mu k^\nu \label{eq:light}
\end{equation}
for all $v$ and the light-like condition $L(v)=0$ is satisfied for the deflected light ray.

\subsection{Area of a perturbed beam} \vspace{-1em}
For each value of the parameter $v$, let $\mathcal{B}_v$ be the two-dimensional surface corresponding to the unperturbed beam. It is coordinatized by  $\sigma^\alpha=(\sigma_1,\sigma_2)$, and its embedding in Minkowski spacetime is given in Cartesian coordinates by $x^\mu_0(v,\sigma)=(v,v,\sigma_1,\sigma_2)$. The corresponding surface for the perturbed beam, which we call $\tilde{\mathcal{B}}_v$, is described by the embedding
\begin{equation}
x^\mu(v,\sigma)=x^\mu_0(v,\sigma)+\xi^\mu(v,\sigma)\,.
\end{equation}
The induced metric on  $\tilde{\mathcal{B}}_v$ is:
\begin{align}
\gamma_{\alpha\beta}(\sigma)&=(\eta_{\mu\nu}+\sqrt{32\pi G}\, h_{\mu\nu}(x(\sigma)))\frac{\partial x^\mu}{\partial \sigma^\alpha}\frac{\partial x^\nu}{\partial \sigma^\beta}\,\nonumber\\
&=\delta_{\alpha\beta}+\partial_\alpha\xi_\beta+\partial_\beta\xi_\alpha+\sqrt{32\pi G}\, h_{\alpha\beta}(x_0(\sigma))\,.
\end{align}
where the dependence of all quantities on the parameter $v$ is understood. $\xi^{\alpha}$ is the projection on the surface of the vector $\xi^\mu$.

The area of the perturbed surface $\tilde{\mathcal{B}}_v$ is 
\begin{equation}
A(v)=\int_{\tilde{\mathcal{B}}_v} \sqrt{\det \gamma_{\alpha\beta}(\sigma)}\,\mathrm{d}^2\sigma\,.
\end{equation}
Expanding the area to the first order in the perturbation we find
\begin{equation}
A(v)=\int_{\mathcal{B}_v} [1+\alpha(v,y)]\,\mathrm{d}^2 y\,,\label{eq:Atilde}
\end{equation}
where the integration is over the unperturbed region $\mathcal{B}_v$ and the perturbation of the area density is defined as
\begin{equation}
\alpha(v,y)\,=\,\partial_\alpha \xi^\alpha(v,y)+\sqrt{8\pi G}\, h^\alpha_\alpha(x_0(v,y))\,.
\end{equation} 
This expressions, together with (\ref{eq:geo}), allows to fully determine the area of a section of the beam. 

It is useful to introduce the notion of expansion $\theta$, defined as the $v$-derivative of the perturbation of the area density:
\begin{equation}\label{deftheta}
\theta = \dot{\alpha}=\partial_\alpha \dot\xi^\alpha+\sqrt{8\pi G}\, \dot h^\alpha_\alpha\,.
\end{equation}
We introduce also the symmetric tensor  $B_{\mu\nu}$
\begin{equation}\label{defB}
B_{\mu\nu}=\frac{1}{2}(\partial_\mu \dot{\xi}_\nu+\partial_\nu \dot{\xi}_\mu)+\sqrt{8\pi G}\, \dot{h}_{\mu\nu}\,.
\end{equation}
Notice that $k^\mu B_{\mu\nu}=k^\nu B_{\mu\nu}=0$ as can be seen using (\ref{eq:geo}) and (\ref{eq:light}). As a result, $B_{\mu\nu}$ is invariant under the projection onto $\mathcal{B}_v$, and the expansion can be expressed as its trace\footnote{The connection with the standard definitions of expansion and shear of a null geodesic congruence in differential geometry is drawn in the Appendix.}, $\theta={B_\mu}^\mu$. Furthermore, using  (\ref{eq:harmonic}) and (\ref{eq:geo}), we obtain a simple expression for the rate of change of the expansion:
\begin{equation}
\dot \theta=\dot {B_\mu}^\mu=\sqrt{8\pi G} \; \Box h_{\mu\nu}(x_0)\, k^\mu k^\nu\,.\label{eq:expansion}
\end{equation}
Using this formula, the area of the perturbed beam can be expressed in terms of the gravitational perturbation.

\subsection{Area of the Rindler horizon} \vspace{-1em}
Consider a family of Rindler observers, labelled by $\ell$, with worldlines
\begin{equation}
x^\mu(\eta)=(\ell \sinh \eta,\,\ell \cosh \eta,\,y_1,\,y_2)\,.\label{eq:Rindler-worldline}
\end{equation}
The Rindler horizon is defined by the boundary of the set of events that can affect a Rindler observer. In the absence of a gravitational perturbation $h_{\mu\nu}$, the Rindler horizon is spanned by the light rays $x_0^\mu(v)$ described by Eq. (\ref{eq:unperturbed}). We are now interested in what happens when a gravitational perturbation is present. 

We assume that the gravitational perturbation vanishes asymptotically: $h_{\mu\nu}(x_0(v)) \to0$ as $v\to\infty$. The light rays that span the Rindler horizon have asymptotically tangent vector $k^\mu$, as the unperturbed light rays. This requirement amounts to the condition
\begin{equation}
 \dot{\xi}^\mu(v)\to 0\,,\quad\quad v\to+\infty\,\label{eq:teleological}
\end{equation}
for the deflection $\xi^\mu$ of the light rays. This final boundary condition is what is often referred to as the teleological nature of the horizon.

We consider now a beam of light rays with initial area $A_0$ at $v=0$, and tuned so that the boundary condition (\ref{eq:teleological}) is satisfied. We compute its asymptotic area $A_H$ in the limit $v\to\infty$. We are interested in the difference  $\Delta A_H = A_H-A^0_H$, where $A^0_H$ is the asymptotic area of the beam in the absence of the gravitational perturbation. The difference $\Delta A_H$ measures the change in area of the Rindler horizon due to the gravitational perturbation $h_{\mu\nu}$.

The asymptotic vanishing of the gravitational field, together with the boundary condition (\ref{eq:teleological}), implies that the expansion vanishes asymptotically. The final boundary condition $\theta(v)\to 0$ can be used to express the expansion $\theta$ in terms of the integral of its derivative:
\begin{equation}\label{advanced1}
\theta(v)=-\int_v^\infty  \dot \theta(v') \,\mathrm{d}v' \,.
\end{equation}
Also integrating (\ref{deftheta}), we can express the perturbation of the area density $\alpha(v)$ in terms of $\dot{\theta}$:
\begin{equation}
\alpha(v)-\alpha(0)=\int_{0}^{v} \theta (v')\mathrm{d}v' 
=-\int_{0}^{v} \mathrm{d}v'' \int_{v''}^\infty \mathrm{d}v' \, \dot \theta(v')\,.\label{eq:alpha-v}
\end{equation}
In the limit $v\to\infty$, using the formula $\int_0^\infty \mathrm{d}v''\int_{v''}^\infty \mathrm{d}v'=\int_0^\infty \mathrm{d}v'\int_0^{v'} \mathrm{d}v''$ for exchanging the integrals in $v'$ and $v''$, we find
\begin{equation}
\alpha(\infty)\,=\,\alpha(0)-\int_{0}^{\infty} \dot{\theta} (v)\,v\mathrm{d}v\,.
\end{equation}
Given the area $A_0$ of the section $\tilde{\mathcal{B}}_0$ of the beam at $v=0$, we can express the asymptotic area $A_H$ of the section $\tilde{\mathcal{B}}_\infty$ as
\begin{equation}
A_H = A_0 -\int_{\mathcal{B}_\infty} \mathrm{d}^2 y \int_{0}^{\infty} \dot{\theta} (v)\,v\mathrm{d}v\,.
\end{equation}
Using (\ref{eq:expansion}), we obtain that the asymptotic area of the beam with the final boundary condition (\ref{eq:teleological}) is
\begin{equation}
A_H \,= A_0+\,\sqrt{8\pi G} \int_{\mathcal{B}_\infty}\!\!\!\mathrm{d}^2 y\int_0^\infty\!\!\! \mathrm{d}v\,v\,[-\Box h_{\mu\nu}(x_0(v))] k^\mu k^\nu\, . \label{areachange0}
\end{equation}
This is the expression we were seeking. Clearly, in the absence of the gravitational perturbation, the asymptotic area of the beam is $A_H^0=A_0$. As a result, the difference in asymptotic areas $\Delta A_H = A_H-A^0_H$ is simply given by
\begin{equation}
\Delta A_H \,=\,\sqrt{8\pi G} \int\mathrm{d}^2 y\int_0^\infty\!\!\! \mathrm{d}v\,v\,[-\Box h_{\mu\nu}(x_0(v))] k^\mu k^\nu\,, \label{areachange1}
\end{equation}
and is independent from the initial area $A_0$ of the beam.

\section{III. The first and the second law} \vspace{-1em}

We now relate the gravitational perturbation $h_{\mu\nu}$ to the energy-momentum tensor $T_{\mu\nu}$ in flat space through the field equations (\ref{fieldequation}), and derive the mechanical laws of the Rindler horizon.

Using the field equations (\ref{fieldequation}) and the fact that $k^\mu$ is light-like, we can express the difference in horizon areas (\ref{areachange1}) as
\begin{equation}\label{areaenergy}
\Delta A_H =8\pi G\!\int\mathrm{d}^2y\int_0^{\infty}\mathrm{d}v\,v\,T_{\mu\nu}(v,y)\,k^\mu k^\nu\,.
\end{equation}
The Second Mechanical Law of the Rindler Horizon follows immediately from this expression. If matter satisfies the Null Energy Condition, $T_{\mu\nu}k^\mu k^\nu\equiv T_{kk}\geq 0$, then it follows immediately that a gravitational perturbation increases the area of the Rindler horizon,
\begin{equation}\label{second}
\Delta A _H = 8\pi G\int\mathrm{d}^2y\int_0^{\infty}\mathrm{d}v\,v\,T_{kk}(v,y)\,\,\,\geq 0\,.
\end{equation}

The First Mechanical Law requires slightly more analysis. It relates the difference in horizon area to the integrated energy flux through the horizon as measured by a Rindler observer. 

The energy-momentum current measured by a Rindler observer is $Q_\mu=T_{\mu\nu} u^{\nu}$. Here $u^\mu=\kappa\, \partial x^\mu/\partial \eta$ is the tangent vector to the world-line (\ref{eq:Rindler-worldline}) of the Rindler observer. The acceleration of the observer is $\kappa$, and the tangent vector is normalized to $-1$ on the world-line with $\ell=\kappa^{-1}$, but defined everywhere as a vector field. In particular, on the future unperturbed Rindler horizon we have 
\begin{equation}
u^\mu= \kappa\,v\, k^\mu\,.
\end{equation}
The integrated energy flux through the unperturbed Rindler horizon is the integral of $Q_\mu$ in the direction $k^\mu$ normal to the horizon
\begin{align}
\Delta Q &= \int\mathrm{d}^2y\int_0^{\infty}\mathrm{d}v\,\, Q_\nu\,k^\nu \nonumber\\
&= \int\mathrm{d}^2y\int_0^{\infty}\mathrm{d}v\,\, \kappa\,v\,T_{\mu\nu}\,k^\mu\,k^\nu\,,\label{eq:DeltaQ}
\end{align}
Using the results derived in the previous section, the energy flux through the Rindler horizon in the presence of a gravitational perturbation can also be computed. In the approximation we are working, the correction can be neglected as of the next order in $G$.

Comparing with (\ref{areaenergy}) we obtain the First Law of Horizon Mechanics:
\begin{equation}\label{firstlaw}
\Delta A_H= \frac{8\pi\,G}{\kappa}\,\Delta Q\,.
\end{equation}
This equation relates the change in horizon area due to the gravitational deflection of the light rays that span it, to the integrated energy flux through the horizon. The derivation of this expression is accurate up to corrections which are higher order in $G$. 

We notice also that, under the assumption that no energy escapes to infinity, the first law can be expressed in terms of the Rindler energy $E_R$. The energy measured at $t=0$ by a Rindler observer with acceleration $\kappa$ is
\begin{equation}
E_R=\int\mathrm{d}^2y\int_0^\infty\!\!\! \mathrm{d}x\; \kappa\, x\, T_{tt}(x,y)\,.\label{eq:E_R}
\end{equation}
As the energy is conserved, the Rindler energy equals the integrated energy-flux $\Delta Q$ through the Rindler horizon plus the energy $E_\infty$ that reaches future null infinity, $E_R=\Delta Q+E_\infty$. If all the energy at $t=0$ and $x>0$ crosses the Rindler horizon, the first law can be stated as  $\Delta A_H= \frac{8\pi\,G}{\kappa}\,E_R$.

\section{IV. Second law: local analysis} \vspace{-1em}
In this section we seek to replace the ``global'' version of the second law, which involves the difference of asymptotic area $A_H - A_H^0$, by a localized version involving the area $A(v)$ of the beam as a function of $v$.

From equation (\ref{eq:alpha-v}) for the evolution of the area density, we find that the evolution of a beam with final boundary conditions (\ref{eq:teleological}) and initial area $A_0$ at $v=0$ is
\begin{equation}
A(v)\,=\,A_0+8\pi G\int_{\mathcal{B}_0} \mathrm{d}^2y \int_0^v\!\!\!\mathrm{d}v''\int_{v''}^\infty\!\!\!\mathrm{d}v'\,T_{kk}(x_0(v',y))\,.
\end{equation}
Its $v$-derivative is manifestly positive as a consequence of the Null Energy Condition $T_{kk}\geq 0$,
\begin{equation}
\dot{A}(v)\,=\,8\pi G\int_{\mathcal{B}_0} \mathrm{d}^2y \int_{v}^\infty\!\!\!\mathrm{d}v'\,T_{kk}(x_0(v',y))\;\geq 0\,.\label{localsecondlaw}
\end{equation}
This proves a local version of the second law: the cross-sectional area of the beam increases with $v$.

Equation (\ref{localsecondlaw}) makes it clear also that there is no similar differential version of the First Law: the infinitesimal increase of the horizon area at a value $v_0$ does not correspond to the local energy flux at $v_0$. In fact, knowledge of  $T_{kk}$ over the whole region from $v_0$ to $\infty$ is needed. This might seem paradoxical but it is a harmless consequence of the teleological nature of the horizon.

It is interesting to analyze the area increase when a localized matter distribution crosses the horizon. Specifically, we assume that $T_{\mu\nu}(x_0(v,y))$ is non-vanishing only in the range $v_0<v< v_1$. From (\ref{localsecondlaw}) it is clear that the area remains constant for $v>v_1$ as the energy flux vanishes. From (\ref{localsecondlaw}) we see also that $\dot{A}(v)$ is constant for $v<v_0$. Therefore there are three regimes in the expansion of a beam spanning the Rindler horizon: (i) an ``inertial'' expansion linear in $v$ in the range $0<v<v_0$, (ii) a ``decelerated'' expansion during the energy flux, $v_0<v<v_1$, (iii) after the passage of matter at $v=v_1$, the beam reaches its asymptotic area and maintains it.

A particularly simple case is the one where the energy flux is completely localized at a time $v_0$. In this case, the area of the beam expands linearly for $v<v_0$, i.e. $A(v)=A_0+(A_H-A_0)v/v_0$, and for $v>v_0$ it remains constant, $A(v)=A_H$.

\section{V. Physical interpretation} \vspace{-1em}

The change in area of the Rindler horizon has a physical interpretation that is well illustrated in terms of deflection of light rays of a laser beam. The wavelength of the beam is assumed to be small enough to make diffraction effects negligible, so an analysis based on geometric optics is valid. 

At $x=0$ and $t=0$, a laser beam with cross-section $A_0$ is fired in the positive $x$-direction. We consider first the case when no gravitational perturbation is present. As the beam is initially collimated, its light rays remain parallel and the cross-section of the beam remains constant and equal to $A_0$. If a gravitational perturbations is present, for instance because some (transparent) matter crosses the beam, then the light rays of the beam are deflected and follow the path (\ref{eq:path}). Because of the attractive nature of gravity, the beam is focused and its cross-sectional area shrinks.

\begin{figure}[htp]
\begin{center}
\includegraphics[width=0.45\textwidth]{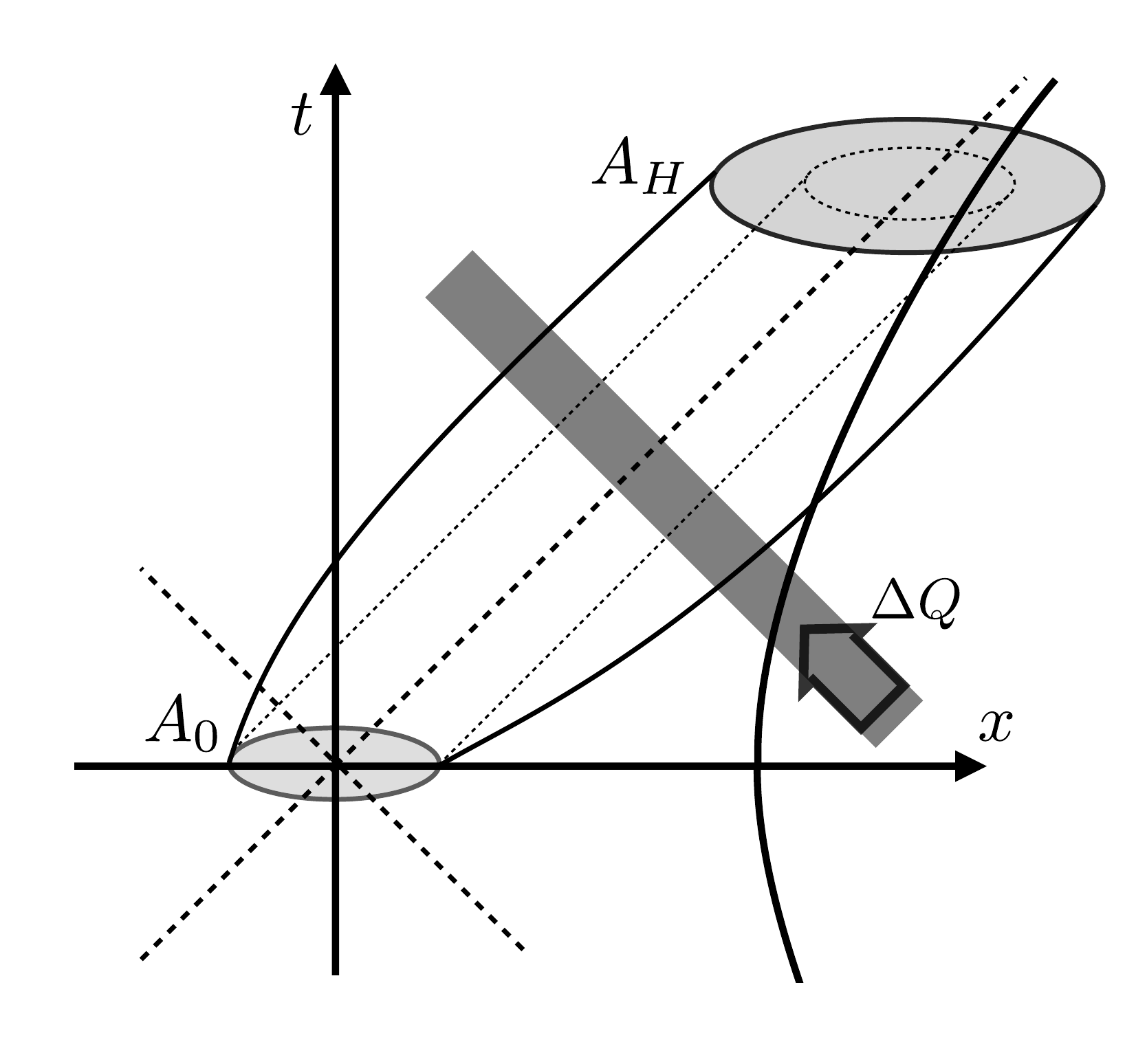}
\end{center}
\caption{\footnotesize{Pictorial illustration of the setting. The shaded discs should be understood as cross sections in the $y_1,y_2$ plane.}}
\label{fig:pictorial}
\end{figure}

\begin{figure}[htp]
\begin{center}
\includegraphics[width=0.45\textwidth]{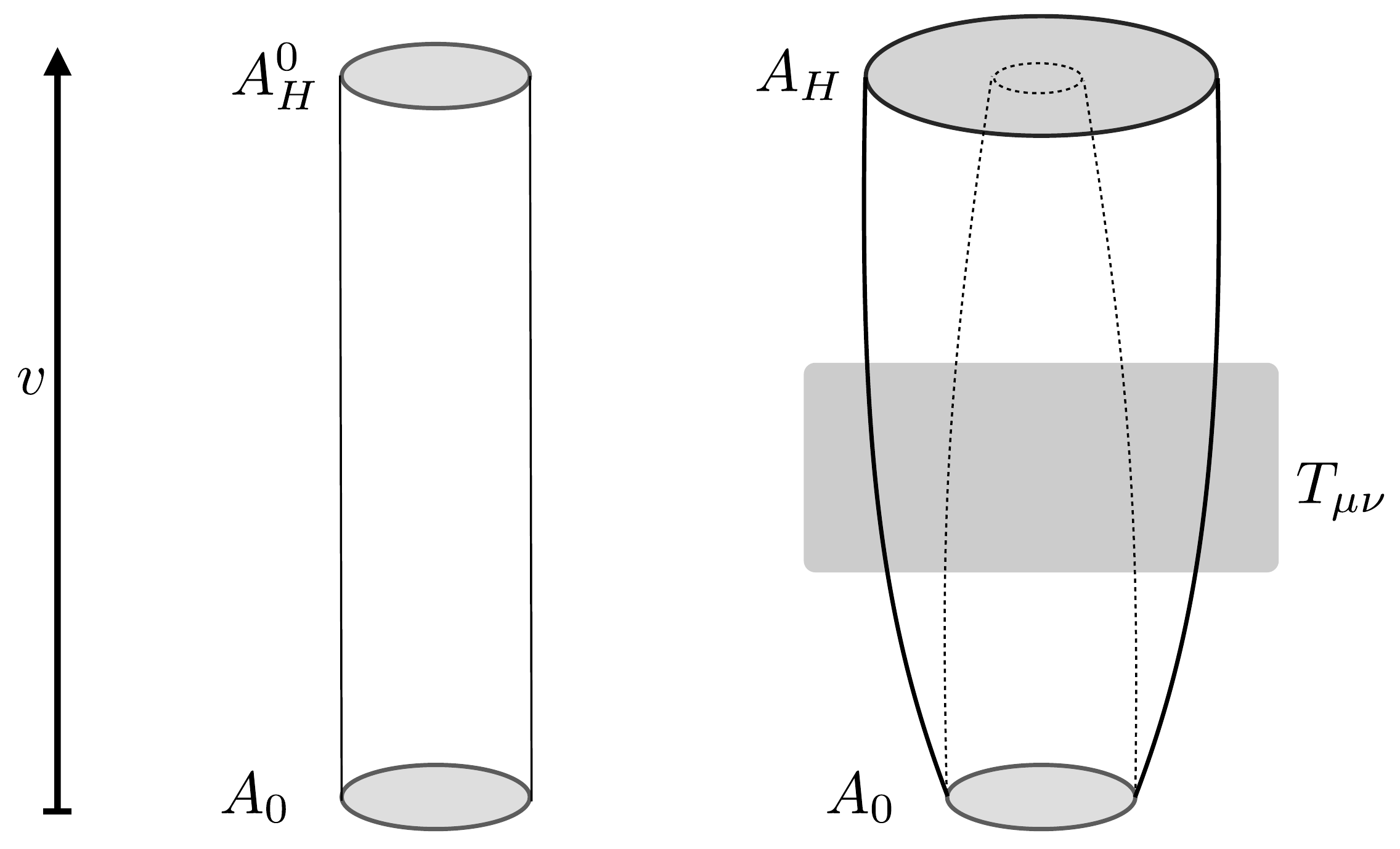}
\end{center}
\caption{\footnotesize{Evolution of a light beam from initial area $A_0$ to asymptotically constant area $A_H$: in the unperturbed case (left panel), an initially collimated beam of area $A_0$ preserves its area; in the presence of a gravitational perturbation (right panel), the collimated beam is focused (dashed lines). To attain an asymptotically constant area $A_H$, the perturbed beam has to be initially expanding.}}
\label{scenarios}
\end{figure}

Now we consider a Rindler observer of the family (\ref{eq:Rindler-worldline}). The associated Rindler horizon consists of light rays that are finely balanced between being detect by the Rindler observer and falling out of sight. This requirement is coded in the final boundary condition (\ref{eq:teleological}). For the beam to reach infinity and be collimated despite the gravitational perturbation, the initial conditions have to be finely tuned: for instance we can place a diverging lens in front of the laser at $t=0$. The initially diverging light rays are focused by the gravitational perturbation and reach infinity collimated. As a result, the cross-sectional area of the beam increases. This situation is depicted in Fig. \ref{fig:pictorial}. The two settings are schematically represented side by side in Fig. \ref{scenarios}.

The First and the Second Laws of Horizon Mechanics both concern the difference $\Delta A_H$ between the asymptotic area $A_H$ of the perturbed beam, and the asymptotic area $A_H^0$ of the unperturbed beam, which equals its original area $A_0$. They relate this difference to the integrated energy flux through the horizon, as measured by the Rindler observer, and state its positivity. Notice that here we are comparing two distinct configurations that are both asymptotically stationary. However, since $\Delta A_H$ also equals the expansion of the perturbed beam from $v=0$ to infinity, the laws can be equally interpreted  as ``physical process'' laws for the temporal evolution of a perturbed horizon.

This twofold interpretation can be compared with the analogous situation for a black hole horizon. There are two logically independent versions of the first law of black hole mechanics \cite{Wald:1995yp}. The first is a physical process version \cite{Hawking:1972hy,1979grec.conf..294C}: one considers an initially stationary black hole; after a small perturbation, the black hole `settles down' to a new stationary state. While the methods used in its proof are very similar to the ones discussed here and in \cite{Jacobson:2003wv,Amsel:2007mh}, there is a key difference: for the Rindler horizon, the initial configuration of the beam is not stationary. In fact at $v=0$ the beam is necessarily diverging, and it becomes collimated and stationary only after the passage of matter.

The second version of the first law of black hole mechanics is an ``equilibrium state comparison'' \cite{Bardeen:1973gs, Sudarsky:1992ty, Wald:1993ki}. In this version one compares two stationary space-times that are infinitesimally close. This is mirrored in our laws holding for the difference between the asymptotic (stationary) areas of two nearby configurations, the perturbed and the unperturbed beam.

In next section we illustrate the physical interpretation of the first law by a concrete example.

\section{VI. A concrete example} \vspace{-1em}

A Rindler observer, Alice, is in a spaceship undergoing uniform acceleration $\kappa$. While preparing a cup of coffee, Alice accidentally drops a bag of sugar, which falls through a trapdoor and leaves the spaceship. The sugar eventually passes through the Rindler horizon. What area change will it induce?

Without loss of generality we can assume that the bag of sugar is dropped when Alice is momentarily at rest in the inertial frame $(t,x,y_1,y_2)$. This happens at $t=0$ and $x=\ell_0=c^2\kappa^{-1}$, or equivalently at $\eta=0$ along her world-line (\ref{eq:Rindler-worldline}). The bag of sugar will remain at rest at $x=\ell_0$ and eventually cross the Alice's horizon at $v=\ell_0$ (we recall that $v$ is defined as $v=\frac{x+ct}{2}$, and we use SI units in this section).

Modeling the package as a point particle, which is valid if its dimensions are much smaller than $l_0$, we have then that the component $T_{kk}$ of its energy-momentum tensor on the horizon is given by
\begin{equation}\label{Tmunuparticle}
T_{kk}= m\, \delta(v-l_0)\,\delta^2(y)
\end{equation}
and hence using (\ref{areaenergy}):
\begin{equation}\label{areasugar}
\Delta A_H = \frac{8\pi\, G\, m\,l_0 }{c^2}=\frac{ 8\pi \,G\,m}{\kappa}\,.
\end{equation}
Note that expressed in terms of $m$ and $\kappa$, the effect is independent from $c$, which suggests that it has an analogue in Newtonian gravity\footnote{Following Mitchell and Laplace, one can introduce a Newtonian version of black hole horizon in terms of the escape velocity $v$ from a star: the horizon area is $A_H=4\pi\, r_H^2$, with $r_H=2GM/v^2$. If we add a mass $m$ to the star, the area of the horizon changes by $\delta A_H=  8\pi \,G\,m/\kappa$, where $\kappa=GM/r_H^2$ is the Newtonian surface gravity, thus reproducing equation (\ref{areasugar}). We thank T.~Jacobson for inspiring
 this observation.}. For the particular values 
\begin{displaymath}
\kappa\sim g_{\mathrm{Earth}}\approx 10\, \mathrm{m/s^2}\,, \quad m\approx 1\, \mathrm{kg}\,,
\end{displaymath}
this amounts to a difference in horizon area
\begin{displaymath}
\Delta A_H \approx 1.7\times 10^ {-10} \mathrm{m}^2\,,
\end{displaymath}
in other words, about 200 $\mu m^2$. This is not an exceedingly small magnitude considered the everyday values of the mass and the acceleration; the reason the effect is undetectable in practice, however, is that  Alice's horizon is approximately one light-year away, $l_0=c^2 g_{\mathrm{Earth}}^{-1}\approx 9\times10^{15} \mathrm{m}$.

It is interesting to consider also the situation where Alice is hovering at a small radial distance $\ell_0$ from a large Schwarzschild black hole. When $\ell_0$ is much smaller than the Schwarzschild radius, the situation is well described by the setting of Rindler mechanics of this paper. In particular, the mass of the black hole drops from the result in this limit. For a bag of sugar left from one meter away from the black hole horizon, the change in horizon area is 
\begin{equation}
\Delta A_H = \frac{8\pi\, G\, m\,l_0 }{c^2}\approx 10^{-26}\mathrm{m}^2\,,
\end{equation}
a finite but small perturbation of the area of the black hole.

\section{VII. Validity of the approximation} \vspace{-1em}

As stated earlier, our calculations are valid at the first order in the gravitational perturbation (or, equivalently, the gravitational coupling $G$) and in the deviation of the light rays, $\xi^\mu$. It is worth stressing that these two assumptions are independent and both needed for our result. It is possible for a weak gravitational perturbation (in the sense that $\sqrt{G} h_{\mu\nu} \ll 1$) to affect light rays in such way that causes a large change in their trajectory -- indeed, this is bound to happen if enough time is given. In this section we investigate under which conditions our truncation of all computations to the lowest order in $\xi^\mu$ is a meaningful and valid one.

Let us return to expression (\ref{areaenergy}) and assume that the matter distribution $T_{\mu\nu}$ is a mass $m$, spread over a region of size $R_0$. As in the example of the previous section, we assume the centre of mass to be at rest at $x=v_0$ in the inertial frame $(t,x,y_1,y_2)$, so that it crosses the Rindler horizon at $v=v_0$. We assume also that
\begin{equation}\label{constraints}
R_0 \leq v_0\,;\quad\quad R_0^2\leq A_0\,.
\end{equation}
The first inequality amounts to requiring that the matter distribution is located entirely at $x>0$, and the second to requiring that the initial cross-section $A_0$ of the beam we track is larger than the cross-section of the perturbation (since otherwise part of the effect is ``missed'').

In this setting a rough estimate of the order of magnitude of the area change is given by:
\begin{equation}\label{estimate}
\Delta A_H \sim G\,m\, v_0\left(1+\mathrm{O}[R_0/v_0]\right)\,,
\end{equation}
as exemplified by (\ref{areasugar}) where we assumed $R_0\ll v_0$.

We need two conditions to be satisfied for our perturbative calculations that yield (\ref{estimate}) to be reliable: weak gravitational perturbation, which in this setting is encompassed by the requirement:
\begin{equation}\label{Schw}
\frac{G m}{R_0}\ll 1\,,
\end{equation}
and small light deflection, which is given by
\begin{equation}\label{deflect}
\Delta A_H \ll A_0\,.
\end{equation}
Condition (\ref{Schw}) implies that the perturbation's spread is much smaller than its Schwarzschild radius and thus the gravitational field is weak, while condition (\ref{deflect}) states that the original area of the beam changes very little and therefore each of the light rays suffers little deflection. 

A sufficient condition for ensuring that both (\ref{Schw}) and (\ref{deflect}) are satisfied is:
\begin{equation}\label{validity}
\frac{G\,m\, v_0}{R_0^2}\ll 1\,.
\end{equation}
This is because this assumption together with the first of the constraints in (\ref{constraints})  implies (\ref{Schw}), and together with the second one implies (\ref{deflect}). Eq. (\ref{validity}) is therefore the desired consistency condition on the validity of the perturbative approximation, relating the mass of the perturbation, its size, and the location at which the horizon is crossed.\footnote{A version of this constraint, equivalent within our assumptions, was given in \cite{Amsel:2007mh} as a requirement to avoid the formation of caustics. See also \cite{Thorne:1986iy,Jacobson:2003wv} for related discussions.}

If the particle is crossing the horizon with nonzero velocity $\beta$ in the $x$ direction, this velocity enters the expression for $T_{\mu\nu}$ used to compute (\ref{Tmunuparticle}). Then the condition required for the validity of the approximation becomes:
\begin{equation}\label{validity2}
\frac{G\,m\, v_0\,\mathrm{e}^{-\varphi}}{R_0^2}\ll 1
\end{equation}
where $\varphi = \tanh^{-1}\beta$ is the rapidity of the particle. Equation (\ref{validity2}) underscores the invariance of our consistency condition under boosts;  a boost with rapidity $\varphi'$ in the $x$-direction, transforms the horizon-crossing parameter $v_0$ into $v_0\,\mathrm{e}^{\varphi'}$, and the rapidity $\varphi$ in $\varphi+\varphi'$. As a result, the consistency condition (\ref{validity2}) remains valid in all frames.

\section{VIII. Horizon entropy} \vspace{-1em}

Our analysis has so far been purely classical. The first immediate advantage of formulating the dynamics of the Rindler horizon in terms of the gravitational field $h_{\mu\nu}(x)$ is that its quantization is immediate: the framework of perturbative quantum gravity as an effective field theory can be used and a quantum version of the first law of the Rindler horizon can be derived.

Let us consider the Minkowski vacuum state $|0\rangle$ and a small perturbation $|\mathcal{E}\rangle$ of the vacuum that corresponds to a non-vanishing energy for a matter field. The integrated energy-flux through the Rindler horizon is defined as
\begin{align}
\Delta Q=\int\mathrm{d}^2y\int_0^{\infty}\!\!\!\mathrm{d}v\,\, \kappa\,v\,\Big(&\langle \mathcal{E}| \hat{T}_{kk}(x_0(v,y))|\mathcal{E}\rangle\,+\\
&\;\;-\langle 0| \hat{T}_{kk}(x_0(v,y))|0\rangle\Big)\,,\nonumber
\end{align}
where the second term in parenthesis corresponds to the standard normal-ordering prescription that assigns zero energy to the vacuum. This is the quantum version of equation (\ref{eq:DeltaQ}). 

The area of the Rindler horizon is defined in terms of the gravitational field $h_{\mu\nu}$. The difference in expectation values of the area on the two states $|0\rangle$ and  $|\mathcal{E}\rangle$ is
\begin{equation}
\Delta A_H=\langle \mathcal{E}|\hat{A}_H|\mathcal{E}\rangle\,-\,\langle 0| \hat{A}_H|0\rangle\,,
\end{equation}
where the operator $\hat{A}_H$ is defined by the quantization of expression (\ref{areachange0}). The gravitational field $h_{\mu\nu}$ consists of a ``transverse'' dynamical part describing gravitons, and a ``longitudinal'' part that is totally constrained by the wave equation (\ref{fieldequation}). At the lowest order in $G$, only this second term contributes and it is immediate to see that equation (\ref{firstlaw}), the first law of the Rindler horizon, holds at the quantum level,
\begin{equation}
\Delta Q\,=\,\frac{\kappa}{8\pi G}\Delta A_H\,.\label{eq:1stlaw-quantum}
\end{equation}
The relation between $\Delta A_H$ and $\Delta Q$ can be studied order by order in perturbation theory to take into account the contribution from gravitons\footnote{We note here that a quantum version of the first law is playing an important role in the microscopic analysis of black hole entropy in loop quantum gravity \cite{Ghosh:2011fc,Frodden:2011eb},\cite{Bianchi:2012ui,Bianchi:2012vp}.}.

In the quantum theory, the Rindler horizon is endowed with a notion of temperature, the Unruh temperature, and a notion of entropy, the entanglement entropy of field-theoretical modes across the horizon. This fact allows to promote the mechanical laws to full-blown thermodynamic laws of the Rindler horizon. The argument, which we summarize here, is explicated at more length in \cite{Bianchi:2012br}. Rindler observers can only measure  observables localized in the Rindler wedge $x>|t|$. As a result, all the expectation values can be computed in terms of a reduced density matrix $\rho$ obtained by tracing over modes outside the Rindler wedge. The entanglement entropy is the Von Neumann entropy of this reduced density matrix, $S[\rho]=-\text{Tr}[\rho\log\rho]$, \cite{Bombelli:1986rw,Srednicki:1993im}. Let us call $\rho(\mathcal{E})$ the reduced density matrix of the state $|\mathcal{E}\rangle$, and $\rho_0$ the reduced density matrix of the vacuum state $|0\rangle$. As is well known, the reduced density matrix of the vacuum state $|0\rangle$ is thermal with respect to the Rindler energy $E_R$ defined in equation (\ref{eq:E_R}),  \cite{Unruh:1976db,Bisognano:1975ih,Bisognano:1976za,Sewell:1982zz}
\begin{equation}
\rho_0\,=\,Z_0^{-1}\,\exp[-\frac{2\pi}{\kappa}\hat{E}_R]\,.
\end{equation}
The local temperature measured by the Rindler observer is $T=\kappa/2\pi$. The excited state $|\mathcal{E}\rangle$ can be considered a small perturbation of the vacuum if the difference $\delta \rho=\rho(\mathcal{E})-\rho_0$ is small (in the sense of all the eigenvalues of $(\delta\rho)^2/\rho_0$ being much smaller than $1$). Working at the first order in the perturbation $\delta \rho$, a Clausius relation for entanglement entropy can be derived
\begin{equation}
\Delta S_{\text{ent}}=S[\rho(\mathcal{E})]-S[\rho_0]=\frac{2\pi}{\kappa}\,\Delta E_R\,
\end{equation}
where $\Delta E_R=\langle \mathcal{E}|\hat{E}_R|\mathcal{E}\rangle\,-\,\langle 0| \hat{E}_R|0\rangle$. Assuming the state $|\mathcal{E}\rangle$ to be such that no energy reaches future null infinity, we have also that $\Delta E_R=\Delta Q$. Then using the quantum version of the first law, equation (\ref{eq:1stlaw-quantum}), we conclude that the difference in entanglement entropy satisfies an area law that reproduces the Bekenstein-Hawking formula,
\begin{equation}
\Delta S_{\text{ent}}=\frac{\Delta A_H}{4 G}\,.
\end{equation}
Therefore, in the quantum theory, under the conditions of small perturbation of the vacuum and no energy escaping to infinity, the first mechanical law of the Rindler horizon is promoted to a thermodynamic law: the acceleration $\kappa$ of the Rindler observer coincides with $2\pi$ times the local Unruh temperature measured by the observer; and the difference in horizon area $\Delta A_H$ between the perturbed and the unperturbed configuration coincides with $4G$ times the difference in the entanglement entropy between the two states.

\section{IX. Summary and conclusions} \vspace{-1em}

In this paper we extended the methods of perturbative gravity in flat Minkowski space to the realm of horizon dynamics: the gravitational field $h_{\mu\nu}$ makes the Rindler horizon dynamical and its area $A_H$ satisfies mechanical laws analogous to the ones satisfied by black holes. Here we briefly summarize our results, their relation with similar results present in the literature, and possible future developments:

1. We determined the area $A_H$ of a portion of the Rindler horizon in terms of the gravitational perturbation $h_{\mu\nu}$, equation (\ref{areachange0}). This expression holds at the lowest order in $G$ and in the deflection $\xi^\mu$. The formula reproduces a standard result in general relativity, where it is derived from the Raychaudhuri equation for the expansion of a null geodesic congruence \cite{Wald:1984rg,poisson2004relativist}. We provided an elementary derivation in terms of the gravitational deflection of light rays in flat space.

2. The difference $\Delta A_H$ between the perturbed and the unperturbed area of the Rindler horizon is related to the energy $\Delta Q$ crossing the horizon, thus providing a derivation of the first mechanical law of the Rindler horizon (\ref{firstlaw}). The derivation is accurate at the first order in $G$, but in principle can be improved order by order using the standard methods of perturbative gravity. These methods can prove useful in providing a perturbative definition of approximate Killing horizons \cite{Guedens:2011dy}.

3. We identified a sufficient condition for the validity of our approximations (\ref{validity2}): the cross-section $R_0^2$ of a matter distribution crossing the Rindler horizon has to be much larger than $G\,m\, v_0\, e^{-\varphi}$, where $m$ is the total mass, $v_0$ the horizon-crossing time, and $\varphi$ the rapidity at the instant of crossing. This condition is compatible with the one found in \cite{Amsel:2007mh} for the absence of caustics.

4. A quantum version of the first law is derived using the methods of perturbative quantum gravity. Together with the Clausius relation for perturbations of the entanglement entropy, this allows to derive a Bekenstein-Hawking formula for the Rindler horizon as done in \cite{Bianchi:2012br}. Similar methods have previously been used to derive the generalized second law of the Rindler horizon \cite{Wall:2010cj,Wall:2011hj}.

The perturbative methods introduced in this paper offer an elementary and reliable tool for studying the mechanics of the Rindler horizon at the classical and at the quantum level. The main interest in these results relies on the idea that the thermodynamics of all horizons has a local nature \cite{Jacobson:2003wv}, and near-horizon processes are well-approximated by the Rindler horizon. From this perspective, insights about the quantum behavior of the Rindler horizon can lead us to a better understanding of the nature of black hole thermodynamics.

\appendix
\section{Acknowledgments} \vspace{-1em}

We thank the Perimeter Institute and the University of Maryland for hospitality during research visits that shaped this paper. We also thank Laurent Freidel, Ted Jacobson, and Flavio Mercati for helpful discussions, and Ted Jacobson for comments on the manuscript. This research was supported by a Banting Fellowship of the NSERC and in part by the National Science Foundation under Grant No. PHY11-25915.

\section{APPENDIX: Raychaudhuri equation and the horizon area} \vspace{-1em}
In this appendix we review the definitions leading to Raychaudhuri's equation for the expansion of a null geodesic congruence \cite{Wald:1984rg,poisson2004relativist}, using the fully geometric language of General Relativity. This leads to a clarification of the relation between our perturbative analysis and  other treatments of the mechanics of approximate Killing horizons \cite{Jacobson:2003wv,Amsel:2007mh}.

Consider a $4$-dimensional Lorentzian space-time $\mathcal{M}$ with coordinates $x^\mu$, metric $g_{\mu\nu}$, and a space-like 2-dimensional surface $\mathcal{B}_0$ with coordinates $y^\alpha$, embedded in $\mathcal{M}$. The surface $\mathcal{B}_0$ possesses two future-pointing null directions and associated to them two families of light rays $F$ and $H$ (null geodesic congruences): the `left-pointing' and the `right-pointing' one. We call the tangent vectors $l^\mu$ and $k^\mu$ respectively,
\begin{equation}
l^\mu=\frac{\partial x^\mu}{\partial u}\,,\;\;k^\mu=\frac{\partial x^\mu}{\partial v}\,,
\end{equation}
where $u$ and $v$ are affine coordinates along light rays in $F$ and $H$. They are null, $k_\mu k^\mu=0,\,l_\mu l^\mu=0$, and on $\mathcal{B}_0$ they satisfy $l_\mu k^\mu\stackrel{\;\mathcal{B}_0\!\!}{=}-2$. The induced metric on $\mathcal{B}_0$ is
\begin{equation}
\gamma_{\mu\nu}=g_{\mu\nu}+\frac{1}{2}(l_\mu k_\nu+l_\nu k_\mu)\,.
\end{equation}
We call $H$ the null geodesic congruence generated by $k^\mu$, and use coordinates $(v,\sigma^\alpha)$ on $H$. The vector $k^\mu$ satisfies the geodesic equation 
\begin{equation}
k^\mu \nabla_\mu k^\nu=0\,.
\end{equation}
The area $A_0$ of the surface $\mathcal{B}_0$,
\begin{equation}
A_0=\int_{\mathcal{B}_0} a\, \mathrm{d}^2\sigma\,,
\end{equation}
is defined as the surface integral of the area density $a$ given by
\begin{equation}\label{areadensity}
a=\sqrt{\gamma_{\mu\nu}\Big(\frac{\partial x^\mu}{\partial \sigma^1}\frac{\partial x^\nu}{\partial \sigma^2}-\frac{\partial x^\mu}{\partial \sigma^2}\frac{\partial x^\nu}{\partial \sigma^1}\Big)}\,.
\end{equation}
We call $\mathcal{B}_v$ the parallel transported surface $\mathcal{B}$ along $H$ with affine parameter $v$, and $a(v)$ its area density. 

The null extrinsic curvature to $H$ is defined as
\begin{equation}\label{defBmunu}
B_{\mu\nu}={\gamma_\mu}^\rho{\gamma_\nu}^\sigma\nabla_\rho k_\sigma\,,
\end{equation}
and can be decomposed in expansion $\theta$, shear $\sigma_{\mu\nu}$, and twist $\omega_{\mu\nu}$,
\begin{align}
\theta=&\,\gamma^{\mu\nu}B_{\mu\nu}\,,\label{deftheta2}\\
\sigma_{\mu\nu}=&\,\frac{1}{2}(B_{\mu\nu}+B_{\nu\mu})-\frac{1}{2}\theta\,\gamma_{\mu\nu}\\
\omega_{\mu\nu}=&\,\frac{1}{2}(B_{\mu\nu}-B_{\nu\mu})\,.
\end{align}
As the family of light rays $H$ is surface orthogonal, the twist $\omega_{\mu\nu}$ vanishes. The expansion $\theta$ describes the logarithmic change in $v$ of the area density of the surface $\mathcal{B}_v$, \begin{equation}
\theta=\frac{1}{a}\frac{da}{d v}\,.
\end{equation}
The Raychaudhuri equation describes the change of the expansion along $H$,
\begin{equation}\label{Raychaudhuri}
\frac{d\theta}{d v}=-\frac{1}{2}\theta^2-\sigma_{\mu\nu}\sigma^{\mu\nu}-R_{\mu\nu}k^\mu k^\nu\;,
\end{equation}
where $R_{\mu\nu}$ is the Ricci tensor.

It is clear that the definitions (\ref{defBmunu}) and (\ref{deftheta2}) for $B_{\mu\nu}$ and $\theta$ match those 
introduced perturbatively in this paper in equations (\ref{deftheta}) and (\ref{defB}). In the standard geometric derivation of the first law \cite{Jacobson:2003wv,Amsel:2007mh}, the terms $\theta^2$ and $\sigma^2$ in (\ref{Raychaudhuri}) are assumed to be negligible and the Ricci tensor $R_{\mu\nu}$ is expressed in terms of the energy-momentum tensor of matter $\tau_{\mu\nu}$ via Einsten equations. Moreover, the notion of approximate boost symmetry is introduced to define a Killing energy. The interplay of these approximations is crucial in the geometric derivation of the first law. On the other hand, in our perturbative derivation there are two clear expansion parameters: the coupling constant $\sqrt{8\pi G}$ of the gravitational field $h_{\mu\nu}$ to light rays, and the deflection $\xi^\mu$ of the light rays from the straight unperturbed path. At the leading order,  the terms corresponding to $\theta^2$ and $\sigma^2$ vanish, the Ricci tensor is given by $R_{\mu\nu} = -\sqrt{8\pi G}\, \Box h_{\mu\nu}$, and the Killing energy is simply given by equation (\ref{eq:E_R}). At this order, the first law is easily derived as shown in the paper. Using the same techniques, the calculation can be done at the next-to-leading order. The framework naturally allows higher curvature Lagrangians, treated via effective field equations. It is possible to extend our results to this case and it would be important to compare this treatment with the analysis of \cite{Guedens:2011dy,Mohd:2013jva}.


\providecommand{\href}[2]{#2}\begingroup\raggedright\endgroup

\end{document}